\documentclass[conference]{IEEEtran}

\usepackage{amsmath, amssymb, bm, cite, epsfig, psfrag}
\usepackage{epstopdf}
\usepackage{graphicx}
\usepackage{algorithm,algpseudocode}
\usepackage{array}
\usepackage[margin=10pt,font=small]{caption}
\usepackage[margin=.7in]{geometry}
\usepackage{bbm}
\usepackage{multirow}
\usepackage[usenames,dvipsnames]{xcolor}

\usepackage{subfigure}
\usepackage{etoolbox}
\usepackage{pbox}
\usepackage[width=0.48\textwidth]{caption}
\usepackage{colortbl}
\usepackage{fancyhdr}
\newtoggle{conference}
\togglefalse{conference} 
\graphicspath{{figures/}}

\def\beq{\begin{equation}}
\def\eeq{\end{equation}}
\def\beqa{\begin{eqnarray}}
\def\eeqa{\end{eqnarray}}
\def\beqan{\begin{eqnarray*}}
\def\eeqan{\end{eqnarray*}}

\def\tm1{t\! - \! 1}
\def\tp1{t\! + \! 1}

\usepackage{lipsum}
\usepackage{fancyhdr}
\usepackage{datetime}

\usepackage{tikz}
\usetikzlibrary{calc}

\begin{document}

\newcommand\blfootnote[1]{%
  \begingroup
  \renewcommand\thefootnote{}\footnote{#1}%
  \addtocounter{footnote}{-1}%
  \endgroup
}

\pagestyle{empty}

\title{Statistical Channel Model with Multi-Frequency and Arbitrary Antenna Beamwidth for Millimeter-Wave Outdoor Communications}

\author{
	\IEEEauthorblockN{Mathew K. Samimi and Theodore S. Rappaport\\ NYU WIRELESS, NYU Tandon School of Engineering\\ mks@nyu.edu, tsr@nyu.edu}
}

\maketitle

\begin{tikzpicture} [remember picture, overlay]
\node at ($(current page.north) + (0,-0.25in)$) {M. K. Samimi, T. S. Rappaport, ``Statistical Channel Model with Multi-Frequency and Arbitrary Antenna Beamwidth};
\node at ($(current page.north) + (0,-0.4in)$) {for Millimeter-Wave Outdoor Communications,'' \textit{in 2015 IEEE Global Communications Conference,}};
\node at ($(current page.north) + (0,-0.55in)$) {\textit{Exhibition \& Industry Forum (GLOBECOM) Workshop}, Dec. 6-10, 2015.};
\end{tikzpicture}

\begin{abstract}
This paper presents a 3-dimensional millimeter-wave statistical channel impulse response model from 28 GHz and 73 GHz ultrawideband propagation measurements~\cite{Rap15_3,Samimi15_2}. An accurate 3GPP-like channel model that supports arbitrary carrier frequency, RF bandwidth, and antenna beamwidth (for both omnidirectional and arbitrary directional antennas), is provided. Time cluster and spatial lobe model parameters are extracted from empirical distributions from field measurements. A step-by-step modeling procedure for generating channel coefficients is shown to agree with statistics from the field measurements, thus confirming that the statistical channel model faithfully recreates spatial and temporal channel impulse responses for use in millimeter-wave 5G air interface designs.
\end{abstract}

 \begin{IEEEkeywords}
28 GHz; 73 GHz; impulse response; spatial spectrum; ray-tracing; multipath; time cluster; spatial lobe; millimeter-wave propagation; statistical channel simulator; SSCM; channel simulator; 5G.
 \end{IEEEkeywords}
\section{Introduction}

Statistical channel models are needed for system-level network simulations and air interface design. Current radio-systems use the 3GPP and WINNER spatial channel models, valid for 1 - 6 GHz, and for RF signal bandwidths up to 100 MHz~\cite{3GPP:1,WinnerII}. The COST 2100 models utilize circular regions (e.g., visibility regions) whose size vary as a user equipment (UE) physically moves and interacts with more or less scattering objects to model realistic links~\cite{Liu12}. This  modeling approach supports temporal and spatial channel correlations over large-scale distances, allowing for smooth channel transitions without discontinuities between closely separated UEs (e.g., spatial consistency)~\cite{Liu12,metis}. 

Recently, new channel modeling frameworks have been developed to characterize millimeter-wave (mmWave) bands. The MiWEBA models use a 60 GHz quasi-deterministic channel model, where strong deterministic components are modeled using Friis' free space path loss equation and path-length geometry, while the properties of weaker clusters of multipaths are generated from measurement-based statistical distributions~\cite{miweba}. The METIS models, in contrast, use a combination of map-based and geometry-based stochastic models to generate channel coefficients~\cite{metis}. 

In this paper, the original statistics from a measurement-based channel impulse response (CIR) model~\cite{Samimi15_2} are extended to include multiple frequencies and arbitrary antenna beamwidths to recreate directional CIRs. A multi-frequency 3-dimensional (3-D) CIR model is presented here, based on 28 GHz and 73 GHz ultrawideband channel measurements in New York City~\cite{Rap13:2, Rap15_3}. The model supports arbitrary carrier frequency, RF signal bandwidth, and antenna beamwidth, following a 3GPP-like modeling approach. The model uses time clusters and spatial lobes to represent the propagation channel~\cite{Samimi15_2}. Note that the 28 GHz and 73 GHz frequency bands for outdoor communications are attractive, as the Federal Communications Commission (FCC) and other governments are about to issue rulemaking, bringing these bands into service~\cite{Rap15_3,MacCartney15_2}. 

The measurements~\cite{Rap15_3} provided over 12,000 measured power delay profiles (PDPs) obtained at unique transmitter (TX) - receiver (RX) pointing angles to construct an accurate statistical spatial channel model (SSCM) at both 28 GHz and 73 GHz separately in non-line of sight (NLOS), and in line of sight (LOS) conditions where the 28 GHz and 73 GHz statistics were combined. The LOS statistics from 28 and 73 GHz were lumped into one data set to extract combined 28 GHz and 73 GHz statistics, motivated by the nearly identical channel statistics for LOS environments at both frequencies. For example, the LOS path loss exponents (PLE) are both nearly identical to free space ($n=2$)~\cite{MacCartney14:2,Rap15_3}, and have virtually the same values of average number of multipath components~\cite{Rap15_3}. Thus, one early finding is that LOS mmWave channels  behave very similarly in terms of path loss, temporal, and spatial properties in~\cite{Rap15_2,MacCartney15_2} as long as there is no oxygen absorption. Thus, it makes sense to generate simplified LOS models that pool data at different frequencies when creating CIRs.

\section{Measurement Description}

28 GHz and 73 GHz wideband propagation measurements were performed at 74 and 36 RX locations, and for three and five distinct TX sites, respectively, with TX-RX distances ranging from 31 m to 425 m on the streets of New York City~\cite{Rap13:2,Rap15_3}. A 400 megachips-per-second broadband sliding correlator channel sounder and highly directional horn antennas were used to recover angle of departure (AOD) and angle of arrival (AOA) statistics. The directional, steerable horn antennas were exhaustively rotated in azimuth and elevation in half-power beamwidth (HPBW) step increments, and many thousands of PDPs were collected at distinct azimuth and elevation unique pointing angles, which provided the means to develop the necessary statistical channel models~\cite{Rap15_3}. Note that impact of sidelobes was minor, with -20 dB sidelobe levels, and a cross-polarization discrimination factor of 21 dB and 25.4 dB for the 28 GHz and 73 GHz outdoor measurements~\cite{Rap15_2}, respectively. The measurement system provided excess delays of the multipath arrivals at the many recorded unique TX-RX pointing angles, and complementary 3-D ray-tracing recreated absolute timing of multipath arrivals from TX to RX~\cite{Samimi15_2}. Table I of~\cite{Rap15_3} gives details about the campaigns and equipment used.

%
%
%
%
%
%
%
%
%

\section{3-D Impulse Response Channel Model}

The SSCM presented here uses temporal clusters and spatial lobes to model the mmWave channel, motivated by observations of the collected New York City measurements~\cite{Rap13:2,Rap15_3}. A temporal cluster represents a group of traveling multipath components arriving closely spaced in time from arbitrary angular directions, while a spatial lobe represents a main directional of arrival where many temporal clusters can arrive at different time delays~\cite{Samimi15_2}. In the 3GPP model, a path is defined as a time-delayed multipath copy of the transmitted signal, that is sub-divided into $M = 20$ subpaths, whose delays are identical to the path delay~\cite{3GPP:1,Calcev07:1}, where each multipath component in the channel is a part of the CIR as shown in~(\ref{IREq}). The WINNER model defines a cluster as a propagation path diffused in space, either or both in delay and angle domains, and is sub-divided into 20 rays, where the two strongest clusters are sub-divided into three sub-clusters with delay offsets of 0 ns, 5 ns, and 10 ns~\cite{WinnerII}. In the COST 2100 model, the time delay of a multipath component is the sum of three delays: the base station (BS)-to-scatterer delay, the mobile station (MS)-to-scatterer delay, and the cluster-link delay~\cite{Liu12}. Ultrawideband PDP measurements at mmWave frequencies, obtained with greater temporal (2.5 ns) and narrower spatial ($7^{\circ},10^{\circ}$) resolution over models in~\cite{3GPP:1,WinnerII,Liu12} indicate that temporal clusters are composed of many intra-cluster subpaths with different random delays, as shown in Fig. 13 of~\cite{Rap15_3}. Thus, the approach presented here offers an improvement over existing models developed for lower frequencies and bandwidths.

The double-directional omnidirectional CIR is commonly used to represent the radio propagation channel between a transmitter and receiver, and can be expressed as in~(\ref{IREq})~\cite{Steinbauer01,Samimi15_2},
\begin{equation}\label{IREq}\begin{split}
h_{omni}(t,\overrightarrow{\mathbf{\Theta}},\overrightarrow{\mathbf{\Phi}}) =& \sum_{n=1}^{N} \sum_{m=1}^{M_n}  a_{m,n} e^{j\varphi_{m,n}} \cdot \delta(t - \tau_{m,n}) \\ 
& \cdot \delta(\overrightarrow{\mathbf{\Theta}}-\overrightarrow{\mathbf{\Theta}}_{m,n}) \cdot \delta(\overrightarrow{\mathbf{\Phi}}-\overrightarrow{\mathbf{\Phi}}_{m,n})
\end{split}\end{equation}

\noindent where $t$ denotes absolute propagation time, $\overrightarrow{\mathbf{\Theta}}=(\theta,\phi)_{TX}$ and $\overrightarrow{\mathbf{\Phi}}=(\theta,\phi)_{RX}$ are the vectors of azimuth/elevation AODs and AOAs, respectively; $N$ and $M_n$ denote the number of time clusters (defined in~\cite{Samimi15_2}), and the number of cluster subpaths, respectively; $a_{m,n}$ is the amplitude of the $m$\textsuperscript{th} subpath belonging to the $n$\textsuperscript{th} time cluster; $\varphi_{m,n}$ and $\tau_{m,n}$ are the phases and propagation time delays, respectively; $\overrightarrow{\mathbf{\Theta}}_{m,n}$ and $\overrightarrow{\mathbf{\Phi}}_{m,n}$ are the azimuth/elevation AODs, and azimuth/elevation AOAs, respectively, of each multipath component.

The statistical channel model presented here also produces the joint AOD-AOA power spectra $P(\overrightarrow{\mathbf{\Theta}},\overrightarrow{\mathbf{\Phi}})$ in 3-D obtained by integrating the magnitude squared of~(\ref{IREq}) over the propagation time dimension,
\begin{align}\label{3DEq1}
P(\overrightarrow{\mathbf{\Theta}},\overrightarrow{\mathbf{\Phi}}) &= \int_{0}^{\infty} |h(t,\overrightarrow{\mathbf{\Theta}},\overrightarrow{\mathbf{\Phi}}) |^2 dt\\ 
\begin{split}P(\overrightarrow{\mathbf{\Theta}},\overrightarrow{\mathbf{\Phi}}) &= \sum_{n=1}^{N} \sum_{m=1}^{M_n} |a_{m,n}|^2  \\ \label{3DEq2}
&\cdot \delta(\overrightarrow{\mathbf{\Theta}}-\overrightarrow{\mathbf{\Theta}}_{m,n}) \cdot \delta(\overrightarrow{\mathbf{\Phi}}-\overrightarrow{\mathbf{\Phi}}_{m,n})
\end{split}
\end{align}

\noindent Current channel models~\cite{3GPP:1,WinnerII,Liu12} use \textit{global} azimuth and elevation spreads to quantify the degree of angular dispersion over the $4\pi$ steradian sphere, using~(\ref{3DEq1}) and the equations in Annex A of~\cite{3GPP:1}. The RMS \textit{lobe} angular spread~\cite{Samimi15_2} is different from the global angular spread, as it only considers the strongest measured lobe directions (as opposed to the entire $4\pi$ steradian power spectrum over space). Typical spatial lobes have absolute  and RMS lobe azimuth spreads of $30^{^\circ}$ and $6^{\circ}$~\cite{Samimi15_2}, respectively, thereby quantifying spatial directionality for realistic multi-element antenna simulations, to emulate beamforming of future directional mmWave systems in the strongest angular directions.

The omnidirectional CIR can further be partitioned to yield \textit{directional} PDPs at a desired TX-RX unique antenna pointing angle, and for arbitrary TX and RX antenna patterns,
\begin{equation}\label{DIREq}\begin{split}
h_{dir}&(t,\overrightarrow{\mathbf{\Theta}_d},\overrightarrow{\mathbf{\Phi}_d}) = \sum_{n=1}^{N} \sum_{m=1}^{M_n}  a_{m,n} e^{j\varphi_{m,n}} \cdot \delta(t - \tau_{m,n}) \\ 
& \cdot g_{TX}(\overrightarrow{\mathbf{\Theta}_d}-\overrightarrow{\mathbf{\Theta}}_{m,n}) \cdot g_{RX}(\overrightarrow{\mathbf{\Phi}_d}-\overrightarrow{\mathbf{\Phi}}_{m,n})
\end{split}\end{equation}

\noindent where ($\overrightarrow{\mathbf{\Theta}_d},\overrightarrow{\mathbf{\Phi}_d})$ are the desired TX-RX antenna pointing angles, $g_{TX}(\overrightarrow{\mathbf{\Theta}})$ and $g_{RX}(\overrightarrow{\mathbf{\Phi}})$ are the arbitrary 3-D (azimuth and elevation) TX and RX complex amplitude antenna patterns of multi-element antenna arrays, respectively. In~(\ref{DIREq}), the TX and RX antenna patterns amplify the power levels of all multipath components lying close to the desired pointing direction, while effectively setting the power levels of multipath components lying far away from the desired pointing direction to 0.

\subsection{Step Procedures for Generating Channel Coefficients}
\label{sec:mmWaveProc}

The step procedure for generating temporal and spatial mmWave channel coefficients is outlined below. In the following steps, $DU$ corresponds to the \textit{discrete uniform} distribution, and the notation $[x]$ denotes the closest integer to $x$. Steps 11 and 12 apply to both AOD and AOA spatial lobes.

\noindent \textit{Step 1: Generate the  TX-RX separation distance d (in 3-D) ranging from 30 - 60 m in LOS, and 60 - 200 m in NLOS (based on our field measurements, and may be modified)}:
\begin{equation}\label{step1}
d \sim U(d_{min},d_{max})
\end{equation}
\noindent where, \[
   \begin{cases}
  d_{min}=30 \text{ m}, d_{max} = 60 \text{ m} ,& LOS\\
  d_{min}=60 \text{ m}, d_{max} = 200 \text{ m} ,& NLOS\\
\end{cases}\]
\noindent To validate our simulation, we used the distance ranges in Step 1, but for standards work other distances are likely to be valid. Users located near BSs (i.e., small TX-RX separation) will be power-controlled in the near field~\cite{MacCartney15_2,Thomas15}.

\noindent \textit{Step 2: Generate the total received omnidirectional power Pr (dBm) at the RX location according to the environment type:}
\begin{align}
&P_r (d)[dBm] = P_t[dBm] - PL(d)[dB]\\
&PL[dB](d) = PL(d_0) + 10\overline{n}\log_{10}\left( \frac{4\pi d}{\lambda} \right)+\chi_{\sigma}\\
&PL(d_0) = 20 \times \log_{10}\left(\frac{4\pi d_0}{\lambda}\right) 
\end{align}
\noindent where $P_t$ is the transmit power in dBm, $d_0 = 1$ m, $\lambda$ is the carrier wavelength, $\overline{n}$ is the path loss exponent (PLE) for omnidirectional TX and RX antennas, given in Table~\ref{tbl:50} for 28 GHz or 73 GHz in both LOS and NLOS environments, and $\chi_{\sigma}$ is the lognormal random variable with 0 dB mean and standard deviation $\sigma$~\cite{Rap15_3}. The $d_0=1$ m close-in (CI) free space reference path loss model is a simple physically-based one-parameter (PLE) model~\cite{Rap15_3,Sun15}, that is more stable across frequencies and environments, than the traditional floating-intercept (FI) least-squares regression equation line~\cite{Rap15_3,Thomas15,MacCartney15_2}. Further, the CI and FI models perform similarly over identical data sets, with differences in standard deviations that are within a fraction of a dB~\cite{Rap15_3,MacCartney15_2,Sun15}. Also, the CI model allows the pooling of LOS power statistics at multiple mmWave frequencies without any change in model coefficients (this is not the case for the FI model). Note that we used $n=2$ to simulate free space propagation in LOS.

\noindent \textit{Step 3: Generate the number of time clusters N and the number of AOD and AOA spatial lobes $(L_{AOD}, L_{AOA})$ at the RX location:}
\begin{align}
&N \sim DU[1,6]\\
& L_{AOD} \sim \min \bigg\{ L_{max},\max \Big\{ 1, \text{Poisson}     \big(    \mu_{AOD}   \big) \Big\} \bigg\}\\
& L_{AOA} \sim \min \bigg\{ L_{max},\max \Big\{ 1, \text{Poisson}     \big(    \mu_{AOA}   \big) \Big\} \bigg\}
\end{align}

\noindent where $L_{max}=5$ is the maximum allowable number of spatial lobes, $\mu_{AOD}$ and $\mu_{AOA}$ are the empirical mean number of AOD and AOA spatial lobes, respectively (see Table~\ref{tbl:softTable}). At 28 GHz in NLOS, the maximum number of time clusters observed was 5, while it was 6 at 73 GHz, using a -10 dB threshold based on work in~\cite{Samimi15_2}. We therefore choose 6 to simplify the model across frequency bands. Note that in~\cite{Samimi15_2}, ($L_{AOD},L_{AOA}$) were conditioned upon $N$, but since subpaths from the same time cluster can arrive and depart from arbitrary directions, the number of spatial lobes is here generalized to be independent of the number of time clusters.

\noindent \textit{Step 4: Generate the number of cluster subpaths (SP) $M_n$ in each time cluster:} 
\begin{align}
&M_n \sim DU[1,30]\hspace{.3cm},\hspace{.3cm} n = 1, 2, ... N
\end{align}

\noindent At 28 GHz in NLOS, the maximum and second to maximum number of cluster subpaths were found to be 53 and 30, respectively, over all locations, while at 73 GHz the maximum was 30 in NLOS, therefore 30 is chosen as the upper bound of the uniform distribution for all frequencies. Subpath components were identified using a peak detection algorithm.

\noindent \textit{Step 5: Generate the intra-cluster subpath excess delays $\rho_{m,n}$}: 
\begin{align}
&\rho_{m,n}(B_{bb}) =\left\{ \frac{1}{B_{bb}}\times (m-1) \right\}^{1+X}\\
&m = 1, 2, ..., M_n \hspace{.3cm},\hspace{.3cm} n = 1, 2, ..., N
\end{align}
\noindent where  $B_{bb}=400$ MHz is the baseband bandwidth of the transmitted PN sequence (but can be modified for different baseband bandwidths less than 400 MHz), and $X$ is uniformly distributed between 0 and $X_{max}$. This step ensures a bandwidth-independent channel model, while reflecting observations that intra-cluster subpath delay intervals tend to increase with delay (through the random variable $X$). The upper bound $X_{max}$ is easily adjustable to field measurements (see Table~\ref{tbl:softTable}).

\noindent \textit{Step 6: Generate the cluster excess delays $\tau_n$:} 
\begin{align}\label{mu_tau}
&\tau^{\prime\prime}_n \sim \text{Exp}(\mu_{\tau})\\
&\Delta \tau_n = \text{sort}(\tau^{\prime\prime}_n)-\min(\tau^{\prime\prime}_n)
\end{align}
\vspace{-.6cm}
\begin{align}
&\tau_n =
\begin{cases}  
      0, &n = 1\\		
    \tau_{n-1}+\rho_{M_{n-1},n-1}+\Delta \tau_n+ 25, &n = 2, ..., N 
\end{cases}
\end{align}

\noindent where $sort()$ orders the delay elements $\tau^{\prime\prime}_n$ from smallest to largest, and where $\mu_{\tau}$ is given in Table~\ref{tbl:softTable}. This step assures no temporal cluster overlap with a 25 ns minimum inter-cluster void interval. The value of 25 ns for minimum inter-cluster void interval was found to match the measured data, and makes sense from a physical standpoint, since multipath components tend to arrive in clusters at different time delays~\cite{Saleh87} over many angular directions, most likely due to the free space air gaps between reflectors (buildings, lampposts, streets, etc). The narrowest streets have a typical spatial width of 8 m (25 ns in propagation delay) in New York City, thus physically describing the regularly observed minimum void interval for arriving energy.

\noindent \textit{Step 7: Generate the time cluster powers $P_n$ (mW):}
\begin{align}\label{eqCluster2}
&P^{\prime}_n = \overline{P}_0 e^{-\frac{\tau_n}{\Gamma}} 10^{\frac{Z_n}{10}}\\ \label{eqCluster}
&P_n = \frac{P^{\prime}_n}{\sum^{k=N}_{k=1} P^{\prime}_k}\times P_r [mW]\\
&Z_n \sim N(0, \sigma_Z )\hspace{.2cm},\hspace{.2cm}n = 1, 2, ... N
\end{align}

\noindent where $\overline{P}_0$ is the average power in the first arriving time cluster, $\Gamma$ is the cluster decay time constant, and $Z_n$ is a lognormal random variable with 0 dB mean and standard deviation $\sigma_Z$ (see  Table~\ref{tbl:softTable}). ~(\ref{eqCluster}) ensures that the sum of cluster powers adds up to the total omnidirectional received power $P_r$. Note that $\overline{P}_0$ cancels out in~(\ref{eqCluster}) using~(\ref{eqCluster2}), but can be used as a secondary statistic to validate the channel model~\cite{Samimi15_2}. The 3GPP, WINNER, COST, and METIS models also parameterize an exponential function over delay, as in~(\ref{eqCluster2}), to estimate mean cluster power levels~\cite{3GPP:1,WinnerII,Liu12,metis}. 

\noindent \textit{Step 8: Generate the cluster subpath powers $\Pi_{m,n}$ (mW)}: 
\begin{align}
&\Pi^{\prime}_{m,n} = \overline{\Pi}_0 e^{-\frac{\rho_{m,n}}{\gamma}}  10^{\frac{U_{m,n}}{10}}\\ \label{eqSP}
&\Pi_{m,n} = \frac{\Pi^{\prime}_{m,n}}{\sum^{k=N}_{k=1} \Pi^{\prime}_{k,n}}\times P_n [mW]\\
& U_{m,n} \sim N(0, \sigma_U)
\end{align}
\noindent where $\overline{\Pi}_0$ is the average power in the first received intra-cluster subpath, $\gamma$ is the subpath decay time constant, and $U_{m,n}$ is a lognormal random variable with 0 dB mean and standard deviation $\sigma_U$ (see  Table~\ref{tbl:softTable}); $m=1, 2, ..., M_n$ and $n=1,2,...,N$. ~(\ref{eqSP}) ensures that the sum of subpath powers adds up to the cluster power. For model validation, the subpath path losses were thresholded at \text{180 dB} (maximum measurable path loss~\cite{Rap15_3}). Note: the measurements have much greater temporal and spatial resolution than previous models. Intra-cluster power levels were observed to fall off exponentially over intra-cluster time delay (see Fig. 4 in~\cite{Samimi15_2}).

\noindent \textit{Step 9: Generate the subpath phases $\varphi_{m,n}$ (rad)}:
\begin{align}
&\varphi_{m,n} \sim U(0,2\pi)
\end{align}

\noindent where $m=1, ..., M_n$ and $n=1,2,...,N$. Different from~\cite{Samimi15_2} where phases are estimated from frequency and delays, here the subpath phases are assumed independently and identically distributed (i.i.d), and uniform between 0 and $2\pi$~\cite{Saleh87} since each subpath may experience a different scattering environment.

\noindent \textit{Step 10: Recover absolute time delays $t_{m,n}$ of cluster subpaths using the TX-RX separation distance $d$ (Step 1):} 
\begin{equation}\label{step10}
t_{m,n} = t_0 + \tau_n + \rho_{m,n} \hspace{.3cm},\hspace{.3cm}t_0 = \frac{d}{c}
\end{equation}

\noindent where $m = 1,2,...M_n, n = 1,2,...N$, and $c = 3\times 10^8 \text{ } m/s$ is the speed of light in free space.

\noindent \textit{Step 11a: Generate the mean AOA and AOD azimuth angles $\theta_i(^{\circ})$ of the 3-D spatial lobes to avoid overlap of lobe angles:}
\begin{align}
&\theta_{i} \sim U(\theta_{min},\theta_{max})\hspace{.3cm},\hspace{.3cm} i=1, 2, ..., L\\
& \theta_{min} = \frac{360(i-1)}{L}, \theta_{max}  = \frac{360i}{L}
\end{align}

\noindent \textit{Step 11b: Generate the mean AOA and AOD elevation angles $\phi_i(^{\circ})$ of the 3-D spatial lobes:}
\begin{align}
&\phi_{i} \sim N(\mu,\sigma), i = 1, 2,..., L. 
\end{align}

\noindent Values of $\phi_i$ are defined with respect to horizon, namely, a positive and negative value indicate a direction above and below horizon, respectively. While the 28 GHz measurements used a fixed 10$^{\circ}$ downtilt at the TX, and considered elevation planes of $0^{\circ}$, and $\pm 20^{\circ}$ at the RX, mmWave transceivers will most likely beamform in the strongest directions, as emulated in the 73 GHz measurements~\cite{MacCartney15_2}. Consequently, the provided elevation angle distributions for all frequency scenarios are extracted from the 73 GHz measurements (see Table~\ref{tbl:softTable}). 

\noindent \textit{Step 12: Generate the AOD angles $(\theta_{m,n,AOD},\phi_{m,n,AOD})$ and AOA angles $(\theta_{m,n,AOA},\phi_{m,n,AOA})$ of each subpath component using the spatial lobe angles found in Step 11:}
\begin{align}
&\theta_{m,n,AOD} = \theta_i+(\Delta \theta_i)_{m,n,AOD}\\
&\phi_{m,n,AOD} = \phi_i +(\Delta \phi_i)_{m,n,AOD}\\ \label{eq37}
&\theta_{m,n,AOA} = \theta_j+(\Delta \theta_j)_{m,n,AOA}\\ \label{eq38}
&\phi_{m,n,AOA} = \phi_j +(\Delta \phi_j)_{m,n,AOA}
\end{align}
\begin{align} \text{where:}\hspace{.2cm}
&i \sim DU[1,L_{AOD}] \hspace{.2cm}, \hspace{.2cm}j \sim DU[1,L_{AOA}] \\ \label{norm}
& (\Delta \theta_i)_{m,n,AOD} \sim N(0,\sigma_{\theta,AOD})\\
&(\Delta \phi_i)_{m,n,AOD} \sim N(0,\sigma_{\phi,AOD})\\ \label{eq42}
& (\Delta \theta_j)_{m,n,AOA} \sim N(0,\sigma_{\theta,AOA})\\ \label{Laplace}
&(\Delta \phi_j)_{m,n,AOA} \sim \text{Laplace}(\sigma_{\phi,AOA})
\end{align}

\noindent This step assigns to each multipath component a single spatial AOD and AOA lobe in a uniform random fashion, in addition to a random angular offset within the spatial lobe with distributions specified in\text{~(\ref{norm}) - (\ref{Laplace})}. Note that the Laplace distribution in~(\ref{Laplace}) provided a better fit to all data across frequencies and environments than a normal distribution. The 3GPP model uses a uniform distribution from $-40^{\circ}$ to $+40^{\circ}$ to generate path azimuth AODs, and for path azimuth AOAs uses a zero-mean normal distribution whose variance is a function of path powers for the UMi scenario~\cite{3GPP:1}. The WINNER models use a wrapped Gaussian distribution that is a function of path powers and delays to generate path AODs and AOAs~\cite{WinnerII}.

\subsection{Implementing the step procedures}

To facilitate the implementation of this SSCM, Table~\ref{tbl:50} and Table~\ref{tbl:softTable} provide the necessary parameters required in Steps 2, 3, 5, 6, 7, 8, 11b, and 12 as a function of the frequency-scenarios considered in this work. 

\begin{table}[b!]
\centering
\caption{Measured path loss exponents and shadow factors~\cite{MacCartney14:2,Rap15_3}, used to generate the omnidirectional received power in Step 2 of Section~\ref{sec:mmWaveProc}. }

\begin{tabular}{|c|c|c|c|}
\hline
\textbf{Step \#} 		& \textbf{Frequency} 		& \textbf{Environment} 	& \textbf{Measured}$\bm{(\overline{n},\sigma)}$  \\ \hline
	
\multirow{4}{*}{Step 2} 	&  \multirow{2}{*}{28 GHz}	& LOS 			& (2.1, 3.6 dB)  \\ \cline{3-4}
				& 					& NLOS 			& (3.4, 9.7 dB)  \\ \cline{2-4}
			 	&  \multirow{2}{*}{73 GHz}	& LOS 			& (2.0, 5.2 dB)  \\ \cline{3-4}
				& 					& NLOS 			& (3.3, 7.6 dB)  \\ \hline
\end{tabular}
\label{tbl:50}
\end{table}

\begin{table*}
\centering
\captionsetup{width=\textwidth}
\caption{Key frequency-dependent parameters that reproduce the measured statistics for the combined 28 - 73 GHz LOS, 28 GHz NLOS, 73 GHz NLOS, and combined 28 - 73 GHz NLOS frequency scenarios.}

\begin{tabular}{|c|c|c|c|c|c|}
\hline
\multirow{2}{*}{\textbf{Step \#}} 	& \multirow{2}{*}{\textbf{Input Parameters}}				& \multicolumn{4}{c|}{\textbf{Frequency Scenario}} \tabularnewline \cline{3-6}
						&											& \textbf{28 - 73 GHz LOS} & \textbf{28 GHz NLOS} 	& \textbf{73 GHz NLOS} & \textbf{28 - 73 GHz NLOS} 	\tabularnewline \hline

Step 3						& $\mu_{AOD},\mu_{AOA}$							& 1.9, 1.8			& 1.6, 1.6			& 1.5, 2.5			& 1.5, 2.1				\tabularnewline \hline

Step 5						& $X_{max}$									& 0.2				& 0.5				& 0.5				&0.5					\tabularnewline \hline

Step 6						& $\mu_{\tau} [ns]$								& 123				& 83				&83				& 83					\tabularnewline \hline

\multirow{1}{*}{Step 7}			& $\Gamma [ns], \sigma_Z [dB]$							& 25.9, 1			& 49.4, 3			& 56.0, 3			& 51.0, 3 			           \tabularnewline \hline

Step 8						& $\gamma [ns], \sigma_U [dB]$							& 16.9, 6			& 16.9, 6			&  15.3, 6			&  15.5, 6				\tabularnewline \hline

\multirow{2}{*}{Step 11b} 		& $\mu_{AOD}[^{\circ}], \sigma_{AOD}[^{\circ}]$				& -12.6, 5.9			& -4.9, 4.5			& -4.9, 4.5			& -4.9, 4.5				\tabularnewline \cline{2-6}
						& $\mu_{AOA}[^{\circ}], \sigma_{AOA}[^{\circ}]$				& 10.8, 5.3			& 3.6, 4.8			&3.6, 4.8			& 3.6, 4.8 				\tabularnewline \hline

\multirow{2}{*}{Step 12} 			& $\sigma_{\theta,AOD}[^{\circ}],\sigma_{\phi,AOD}[^{\circ}]$		& 8.5, 2.5			& 9.0, 2.5			& 7.0, 3.5			& 11.0, 3.0				\tabularnewline \cline{2-6}
						& $\sigma_{\theta,AOA}[^{\circ}],\sigma_{\phi,AOA}[^{\circ}]$	 	& 10.5, 11.5			& 10.1, 10.5			& 6.0, 3.5			& 7.5, 6.0					\tabularnewline \hline

\end{tabular}
\label{tbl:softTable}
\end{table*}

\subsection{Sample Output Functions}

\begin{figure}[t]
    \begin{center}
        \includegraphics[width=3.5in]{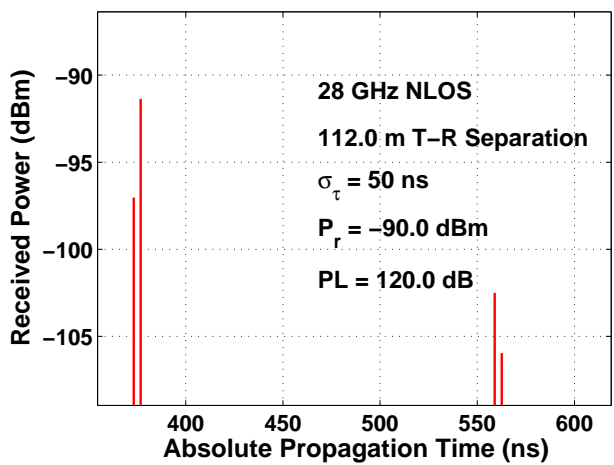}
    \end{center}
   \caption{Example of simulated 28 GHz NLOS PDP, showing four multipaths, obtained from the MATLAB-based statistical simulator, assuming 0 dBi TX and RX antenna gains. Temporal cluster powers and intra-cluster subpath powers decay with increasing delay according to empirical time decay constants, $\Gamma = 49.4$ ns and $\gamma=16.9$ ns, respectively.}
    \vspace{-0.4cm}
    \label{fig:SamplePDP_28GHzNLOS}
    \end{figure}

\begin{figure}[t]
    \begin{center}
        \includegraphics[width=3.5in]{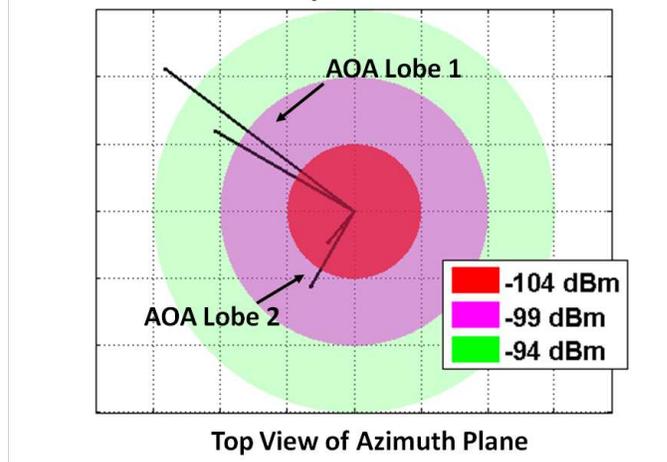}
    \end{center}
    \caption{Example of simulated 28 GHz NLOS 3-D AOA power spectrum (top view of the azimuth plane) of Fig.~\ref{fig:SamplePDP_28GHzNLOS} obtained from the MATLAB-based statistical simulator, showing four multipath components grouped into two AOA spatial lobes.}
    \vspace{-0.4cm}
    \label{fig:SampleAOD_28GHzNLOS}
    \end{figure}

Figs.~\ref{fig:SamplePDP_28GHzNLOS} and \ref{fig:SampleAOD_28GHzNLOS} show output functions of a 28 GHz NLOS omnidirectional PDP, and corresponding AOA 3-D power spectrum, obtained from a MATLAB-based statistical simulator that implemented the channel models given \text{in~(\ref{step1}) - (\ref{Laplace})}. The generated PDP in Fig.~\ref{fig:SamplePDP_28GHzNLOS} is composed of four multipath taps, grouped into two time clusters with exponentially decaying amplitudes with cluster decay constant $\Gamma=49.4$ ns and intra-cluster subpath decay constant $\gamma=16.9$ ns (see Section~\ref{sec:mmWaveProc}, Steps 6 and 7). Here, the simulated PDP has a total path loss of 120 dB with TX-RX separation distance of 112 m, and RMS delay spread of 50 ns, with 0 dBi TX and RX antenna gains. The AOA spectrum (Fig.~\ref{fig:SampleAOD_28GHzNLOS}) shows the four multipath grouped into two AOA spatial lobes according to~(\ref{eq37}),~(\ref{eq38}),~(\ref{eq42}), and~(\ref{Laplace}).

\section{Simulation Results}

The statistical channel models presented in~(\ref{step1}) - (\ref{Laplace}) were implemented in a MATLAB-based statistical simulator to confirm the accuracy of the SSCM against the measured statistics. A large simulation was carried out in which 10,000 omnidirectional PDPs for omnidirectional TX and RX, and 3-D AOD and AOA power spectra were generated according to~(\ref{IREq}) and~(\ref{3DEq1}). Simple number generators were utilized to obtain the number of time clusters, the number of AOD and AOA spatial lobes, cluster and subpath delays, and cluster and subpath powers, as described in Section~\ref{sec:mmWaveProc}. To remain faithful to the measurements, the total dynamic range was set to 180 dB, and the TX and RX 3-dB antenna beamwidths were set to 10$^{\circ}$ and 7$^{\circ}$ (in azimuth and elevation)~\cite{MacCartney15_2}, when performing directional simulations.

\subsection{Simulated RMS Delay Spreads}

\begin{figure}
    \begin{center}
        \includegraphics[width=3.5in]{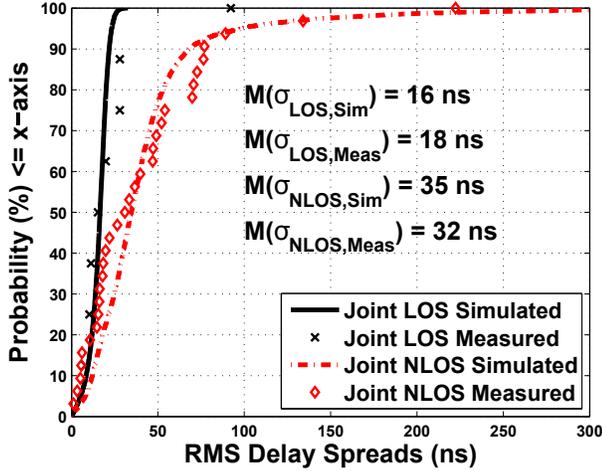}
    \end{center}
    \caption{Combined 28 - 73 GHz LOS and NLOS omnidirectional RMS delay spreads synthesized from absolute timing PDPs, superimposed with 10,000 simulated RMS delay spreads from generated omnidirectional PDPs~\cite{Samimi15_2}. }
    \vspace{-0.4cm}
    \label{fig:JointOmni_DS}
    \end{figure}

Fig.~\ref{fig:JointOmni_DS} compares the omnidirectional simulated RMS delay spreads and empirical values from omnidirectional PDPs~\cite{Samimi15_2}, at both 28 GHz and 73 GHz in LOS and NLOS scenarios.  The empirical and simulated medians were 18 ns and 16 ns, respectively, for the combined 28-73 GHz LOS scenario, and 32 ns and 35 ns for the empirical and simulated medians, respectively, for the combined 28-73 GHz NLOS scenario. The few measured data samples considerably skewed the empirical distributions, so the median (instead of the mean) was selected to represent the distribution trend. The empirical and simulated medians in NLOS were 31 ns and 32 ns at 28 GHz (see Fig. 5 in~\cite{Samimi15_2}), respectively, and 47 ns and 39 ns at 73 GHz, respectively, providing good agreement to empirical values.
   
The model also produces \textit{directional} PDPs at arbitrary TX-RX pointing angle combination, and reconstructs the temporal statistics of arbitrary antenna beamwidths. This is achieved by weighting the multipath component power levels with a user-defined complex amplitude antenna pattern, such that the multipath components closest to a desired direction are amplified, while those farthest away are effectively set to 0, as shown in~(\ref{DIREq}). Using~(\ref{DIREq}), the multipath component power levels were weighted by $|g_{TX}(\theta,\phi)|^2$ and $|g_{RX}(\theta,\phi)|^2$, the TX and RX horn antenna patterns, respectively, commonly parameterized as follows~\cite{Balanis05},
\begin{align}
&|g(\theta,\phi)|^2 = \max \big(   G_0 e^{\alpha \theta^2+\beta \phi^2}, \frac{G_0}{100} \big)   \\
& \alpha = \frac{4 \ln(2)}{\theta_{3dB}^2}, \beta = \frac{4 \ln(2)}{\phi_{3dB}^2}, G_0 = \frac{41253 \eta }{\theta_{3dB}\phi_{3dB}}
\end{align} 

\noindent where $(\theta,\phi)$ are the azimuth and elevation angle offsets from the boresight direction in degrees, $G_0$ is the maximum directive gain (boresight gain) in linear units, $(\theta_{3dB},\phi_{3dB})$ are the azimuth and elevation HPBWs in degrees, $\alpha,\beta$ are parameters that depend on the HPBW values, and $\eta = 0.7$ is a typical average antenna efficiency.

\begin{figure}
    \begin{center}
        \includegraphics[width=3.5in]{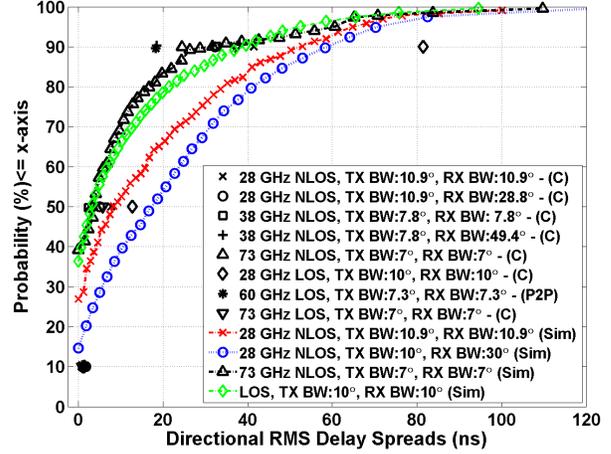}
    \end{center}
    \caption{Simulated directional RMS delay spread CDFs (lines) for various frequencies and for antenna beamwidths, obtained from directional PDPs generated using (\ref{DIREq}). Values reported from the literature are shown as points~\cite{Rap13:2,MacCartney14:2}. In the figure legend, 'C' stands for 'Cellular', 'P2P' stands for Peer-to-Peer', and 'Sim' stands for 'Simulated'.}
    \vspace{-0.4cm}
    \label{fig:RMSDS_Sim}
    \end{figure}

Fig.~\ref{fig:RMSDS_Sim} shows simulated directional RMS delay spreads obtained from the 28 GHz and 73 GHz channel models presented here, obtained from~(\ref{DIREq}), in comparison to reported values in the literature for the 10\%, 50\%, and 90\% CDF points of measured directional RMS delay spreads at 28 GHz, 38 GHz, 60 GHz, and 73 GHz~\cite{MacCartney15_2}, for antenna beamwidths of 7.3$^{\circ}$, 10.9$^{\circ}$, 28.8$^{\circ}$, and 49.4$^{\circ}$. To test~(\ref{DIREq}), we generated sample functions and computed directional RMS delay spreads from directional PDPs for 20 random TX-RX separation distances using antenna HPBW of 10$^{\circ}$, $7^{\circ}$, and $30^{\circ}$ in azimuth/elevation at 28 GHz and 73 GHz to emulate the NYC measurements. Note that the empirical CDFs consist of all available data, while the simulated directional RMS delay spreads were generated from channel models obtained exclusively from up to four strongest AOD and AOA PDP data. Fig.~\ref{fig:RMSDS_Sim} indicates that the simulated and measured omnidirectional and directional RMS delay spread distributions match relatively well across antenna beamwidths and many mmWave bands.

\subsection{Simulated RMS Angular Spreads}

\begin{figure}
    \begin{center}
        \includegraphics[width=3.5in]{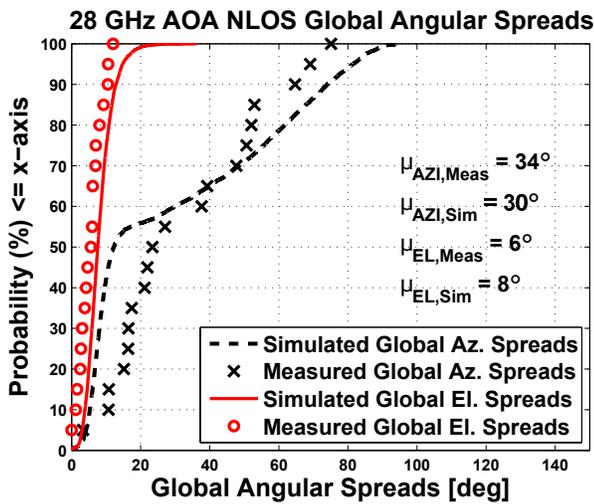}
    \end{center}
    \caption{28 GHz NLOS global azimuth and elevation spreads, obtained from the NYC measurements~\cite{Rap15_3} and the 3-D SSCM presented here.}
    \vspace{-0.4cm}
    \label{fig:GlobalAngularSpreads_AOA_28NLOS}
    \end{figure}

\begin{figure}
    \begin{center}
        \includegraphics[width=3.5in]{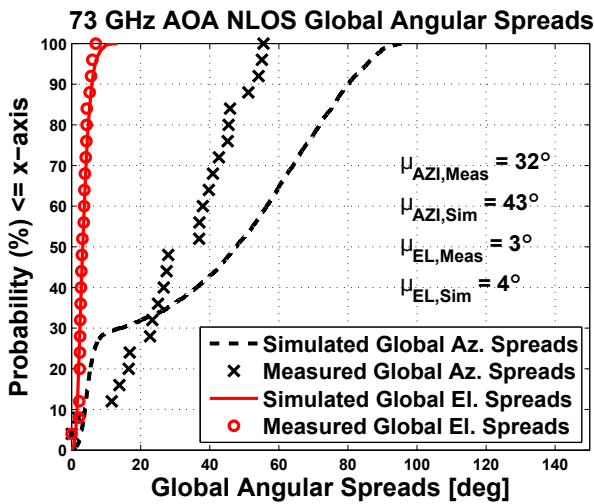}
    \end{center}
    \caption{73 GHz NLOS global azimuth and elevation spreads, obtained from the NYC measurements~\cite{Rap15_3} and the 3-D SSCM presented here.}
    \vspace{-0.4cm}
    \label{fig:GlobalAngularSpreads_AOA_73NLOS}
    \end{figure}

The omnidirectional azimuth and elevation spreads describe the degree of angular dispersion at a BS or MS over the entire $4\pi$ steradian sphere~\cite{3GPP:1,WinnerII}, also termed \textit{global} angular spreads in~\cite{Liu12}. The AOD and AOA global angular spreads were computed from all available 28 GHz and 73 GHz NLOS measured data, using the total (integrated over delay) received power at unique azimuth/elevation pointing angles, but not requiring absolute multipath time delays, and the equations in Annex A of~\cite{3GPP:1}. These were compared with the simulated angular spreads using the 3-D SSCM, where the SSCM was developed from the statistics of up to four strong measured angles. The simulated and measured mean global angular spreads match relatively well at 28 GHz and 73 GHz, as can be seen from Fig.~\ref{fig:GlobalAngularSpreads_AOA_28NLOS} and Fig.~\ref{fig:GlobalAngularSpreads_AOA_73NLOS}. The slight differences (slight under- or over-estimation of global azimuth spreads using Step 12 in Section~\ref{sec:mmWaveProc}) may be due to the model focusing only on the multipath components contained in the strongest several spatial lobes measured at every location, which will in actuality be the strongest components in a practical wireless system.

\begin{figure}
    \begin{center}
        \includegraphics[width=3.5in]{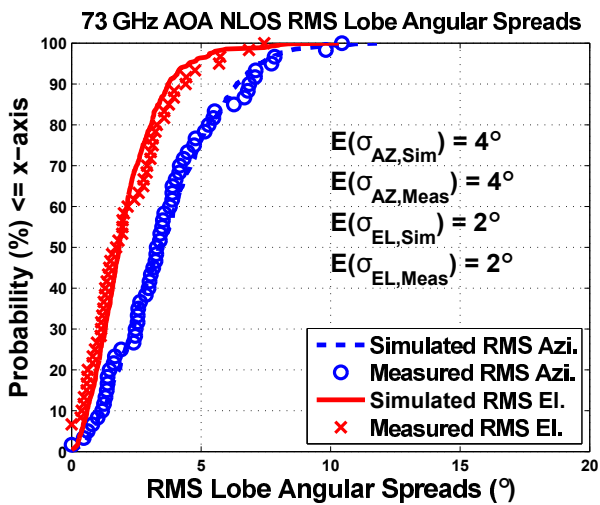}
    \end{center}
    \caption{73 GHz NLOS AOA RMS lobe azimuth and elevation spreads, measured and simulated, showing good agreement~\cite{Samimi15_2}.}
    \vspace{-0.4cm}
    \label{fig:AngularSpreads_AOA_73NLOS}
    \end{figure}

The directional AOD and AOA RMS \textit{lobe} azimuth and elevation spreads were also computed based on a -10 dB lobe threshold~\cite{Samimi15_2} from the 28 GHz and 73 GHz data, and compared with simulated values using the 3-D SSCM. Fig.~\ref{fig:AngularSpreads_AOA_73NLOS} shows typical measured against simulated AOA RMS lobe angular spreads for the 73 GHz NLOS scenario, showing an excellent match over empirical and simulated means of 4$^{\circ}$ and $2^{\circ}$ in azimuth and elevation, respectively, over all measured spatial lobes. Similar agreement was found for the 28 GHz NLOS and LOS datasets, across AOD and AOA spatial lobes, indicating that the model accurately recreates the empirical spatial statistics. 


%
%
%
%
%
%
%
%
%
%
%

\section{Conclusion}

This paper presented a 3-D statistical  model of the channel impulse response for mmWave LOS and NLOS mobile access communications, that supports arbitrary carrier frequency, signal bandwidth, and antenna beamwidth, with good agreement between the model and published directional RMS delay spreads. The RMS \textit{lobe} angular spreads~\cite{Samimi15_2} provide a realistic representation of directional angular spreads for future multi-antenna mmWave systems, where the strongest multipath components contained in spatial lobes are most important for characterizing system behavior. A MATLAB-based statistical simulator was implemented to generate a large ensemble of PDPs and 3-D power spectra, showing good agreement to field measurements, allowing for future millimeter-wave system and bit-error rate simulations, as well 5G wireless network capacity analyses.

\bibliographystyle{IEEEtran}
\bibliography{bibliography_MSThesis}
\end{document}